\documentclass[twocolumn]{webofc}
\usepackage[varg]{txfonts}   
\begin{document}
\title{Transverse single spin asymmetry for very forward $\pi^{0}$ production in polarized proton-proton collisions at $\sqrt{s}$ = 510 GeV}
%
%

\author{\lastname{M. H. Kim}\inst{1,2}\fnsep\thanks{\email{jipangie@korea.ac.kr}} \and
        \lastname{B. Hong}\inst{1} \and
        \lastname{Y. Goto}\inst{2,3} \and
        \lastname{I. Nakagawa}\inst{2,3} \and
        \lastname{R. Seidl}\inst{2,3} \and
        \lastname{J. S. Park}\inst{2,4} \and
        \lastname{K. Tanida}\inst{5} \and
        \lastname{Y. Itow}\inst{6,7} \and
        \lastname{H. Menjo}\inst{10} \and
        \lastname{K. Sato}\inst{6} \and
        \lastname{M. Ueno}\inst{6} \and
        \lastname{Q. D. Zhou}\inst{6} \and
        \lastname{T. Sako}\inst{8} \and
        \lastname{K. Kasahara}\inst{9} \and
        \lastname{T. Suzuki}\inst{9} \and
        \lastname{S. Torii}\inst{9} \and
        \lastname{N. Sakurai}\inst{11} \and
        \lastname{O. Adriani}\inst{12,15} \and
        \lastname{L. Bonechi}\inst{12} \and
        \lastname{E. Berti}\inst{12,15} \and
        \lastname{R. D'Alessandro}\inst{12,15} \and
        \lastname{A. Tricomi}\inst{13,14}
}

\institute{Department of Physics, Korea University, Seoul, South Korea \and 
RIKEN Nishina Center for Accelerator-Based Science, Saitama, Japan \and
RIKEN BNL Research Center, Brookhaven National Laboratory, New York, USA \and
Department of Physics, Seoul National University, Seoul, South Korea \and
Japan Atomic Energy Agency, Ibaraki, Japan \and
Institute for Space-Earth Environmental Research, Nagoya University, Nagoya, Japan \and
Kobayashi-Maskawa Institute for the Origin of Particles and the Universe, Nagoya University, Nagoya, Japan \and
Institute for Cosmic Ray Research, University of Tokyo, Kashiwa, Chiba, Japan \and
RISE, Waseda University, Shinjuku, Tokyo, Japan \and
Graduate School of Science, Nagoya University, Nagoya, Japan \and
Tokushima University, Tokushima, Japan \and
INFN Section of Florence, Florence, Italy \and
INFN Section of Catania, Italy \and
University of Catania, Catania, Italy \and
University of Florence, Florence, Italy
}

\abstract{%
Transverse single spin asymmetry, $A_{N}$, of very forward $\pi^{0}$ production from polarized $p + p$ collisions provides new information toward an understanding of its production mechanism. $A_{N}$ of forward $\pi^{0}$ in the pseudorapidity region of $3 < \eta < 4$ has been described by the partonic structure of the proton in the perturbative QCD framework. However, recent data indicates a potential contribution from not only partonic but also  diffractive interactions. In order to provide a new insight on the origin of the $A_{N}$, we measured the very forward $\pi^{0}$ production in the pseudorapidity region of $6 < \eta$ from $\sqrt{s}$ = 510 GeV polarized $p + p$ collisions at RHIC in 2017. We report our measurement of the very forward $\pi^{0}$ over the transverse momentum range of $0 < p_{T} < 1$ GeV/$c$ and the preliminary result. 
}
\maketitle
\section{Introduction}
\label{sec:intro}
A new experiment, RHIC forward \cite{rhicfproposal} (RHICf), has measured the transverse single spin asymmetry, $A_{N}$, of very forward particle production in $\sqrt{s}$ = 510 GeV polarized proton-proton collisions at the Relativistic Heavy Ion Collider (RHIC) in June, 2017. $A_{N}$ is defined as a left-right asymmetry of the production cross section to beam polarization. It plays an important role in the study of the production mechanism, particularly from the view points of diffractive and non-diffractive interactions. 

To date, non-zero $A_{N}$ in forward $\pi^{0}$ production has been measured by many experiments \cite{fnal, brahms, phenix, star}. Non-zero $A_{N}$ has been observed in only the forward-rapidity region, and it has been interpreted based on the perturbative QCD framework. Typically there are two interpretations. The first one relies on the transverse momentum dependent parton distributions inside the polarized proton and corresponding spin dependent fragmentation known as the Collins fragmentation function \cite{collins}. The other one introduces spin-dependent twist-three quark-gluon correlations \cite{twist1, twist2, twist3} and twist-3 fragmentation functions \cite{twist4}. 

However, recent studies \cite{pi0star1, pi0star2} at the STAR experiment provided a new possibility of a finite contribution of the diffractive interaction to the non-zero $A_{N}$ of forward $\pi^{0}$. Larger $A_{N}$ was observed with more isolated $\pi^{0}$ production and it also decreased when the event complexity increased which was more hard scattering-like event topology. This result shows an agreement with the preliminary observations from the $A_{N}DY$ collaboration \cite{andy}, which showed very small magnitude of $A_{N}$ for the inclusive jet production. In Ref. \cite{andyex}, it is argued that the small jet $A_{N}$ is cancelled out because the asymmetries of $u$ and $d$ quarks have opposite sign but equal magnitude as shown in previous measurements of the $\pi^{\pm}$. However, no consensus has been reached yet for the production mechanism of the forward $\pi^{0}$ asymmetry. One of the best ways to address  this open question is to examine if there is a finite asymmetry in a particular interaction by specifying an event type. 

The RHICf experiment can study the production mechanism of the $\pi^{0}$ production by measuring the very forward $(6 < \eta)$ $\pi^{0}$ in wide transverse-momentum coverage ($0 < p_{T} < 1$ GeV/$c$) where the diffractive process events are enhanced. In this paper, we present our measurement of very forward $\pi^{0}$ in Sec. \ref{sec:RHICf} and analysis procedure in Sec. \ref{sec:analysis}. Our preliminary result is presented in Sec.~\ref{sec:result}.

\section{RHICf experiment}
\label{sec:RHICf}
\begin{figure}[!t]
\centering
\sidecaption
\includegraphics[width=0.47\textwidth]{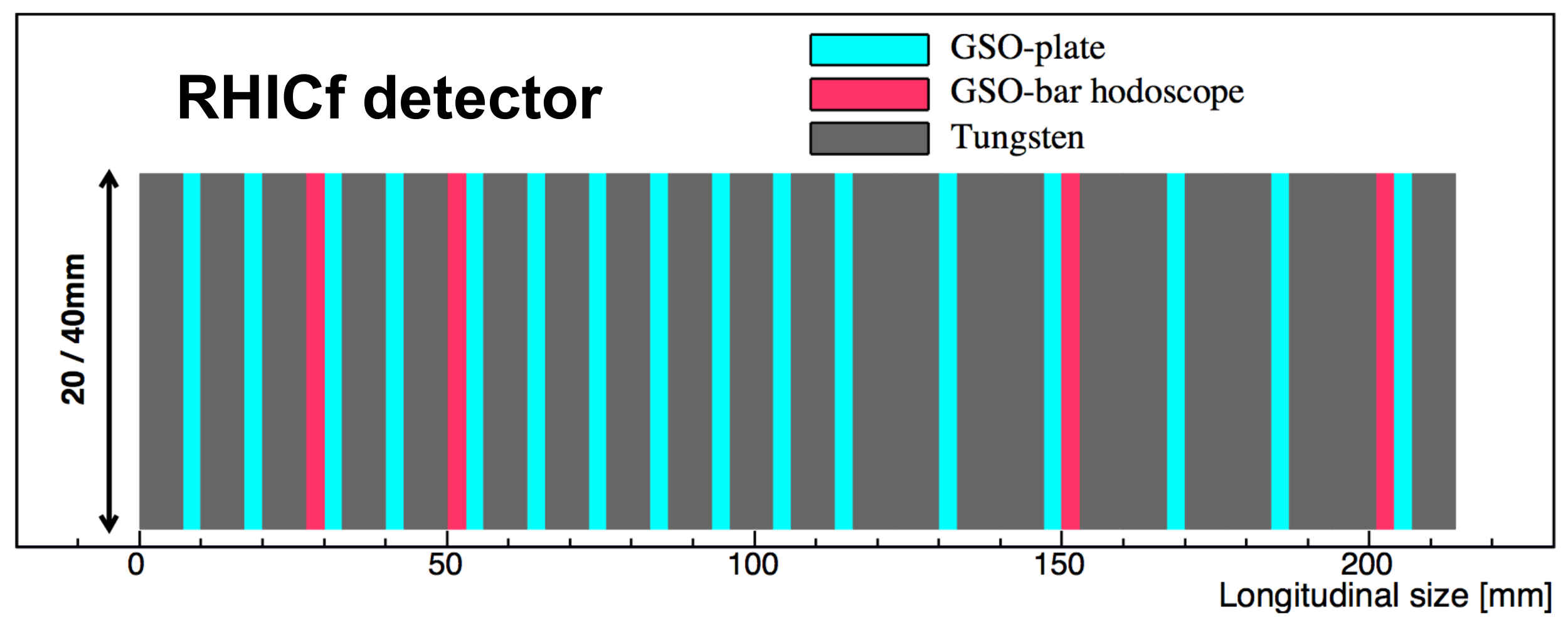}
\caption{Schematic side view of RHICf detector. The structures of small and large towers are same.} \label{fig:detector}
\end{figure}

In 2017, we installed a new electromagnetic calorimeter (RHICf detector) at the STAR experiment in front of one of the hadronic calorimeter ZDC \cite{zdc} which was located 18 m away from the beam interaction point and measured very forward particles, mainly neutron, $\pi^{0}$, and photon. The detector had been originally developed for the LHCf experiment \cite{lhcf} at CERN. Fig. \ref{fig:detector} shows a schematic side view of the RHICf detector. The RHICf detector consists of small and large towers with 20 and 40 mm transverse dimensions respectively.
Each tower is composed of 17 layers of tungsten plates, 16 layers of GSO plates for energy measurement, and 4 layers of thin GSO bars for position measurements. Each GSO bar layer covers the detector with an array of 1 mm wide GSO scintillators \cite{detector}. The longitudinal thickness of the calorimeter is 44 radiation lengths. Electromagnetic showering usually stops its development in the middle of the detector. 

In operation, we took the data with three different detector positions to cover the wide $p_{T}$ range up to 1 GeV/$c$. 110 M events were accumulated for neutral particles during 28 hours (4 Fills) with larger $\beta^{*}$ = 8 m for a small angular beam divergence and lower instantaneous luminosity of $\sim 10^{31}$ cm$^{-2}$s$^{-1}$ than for usual RHIC operation to avoid the multiple collisions and radial beam polarization which was normal to the usual vertical one to cover the wide $p_{T}$ range of neutral particles by moving the detector.

\begin{figure}[ht]
\centering
\sidecaption
\includegraphics[width=0.5\textwidth]{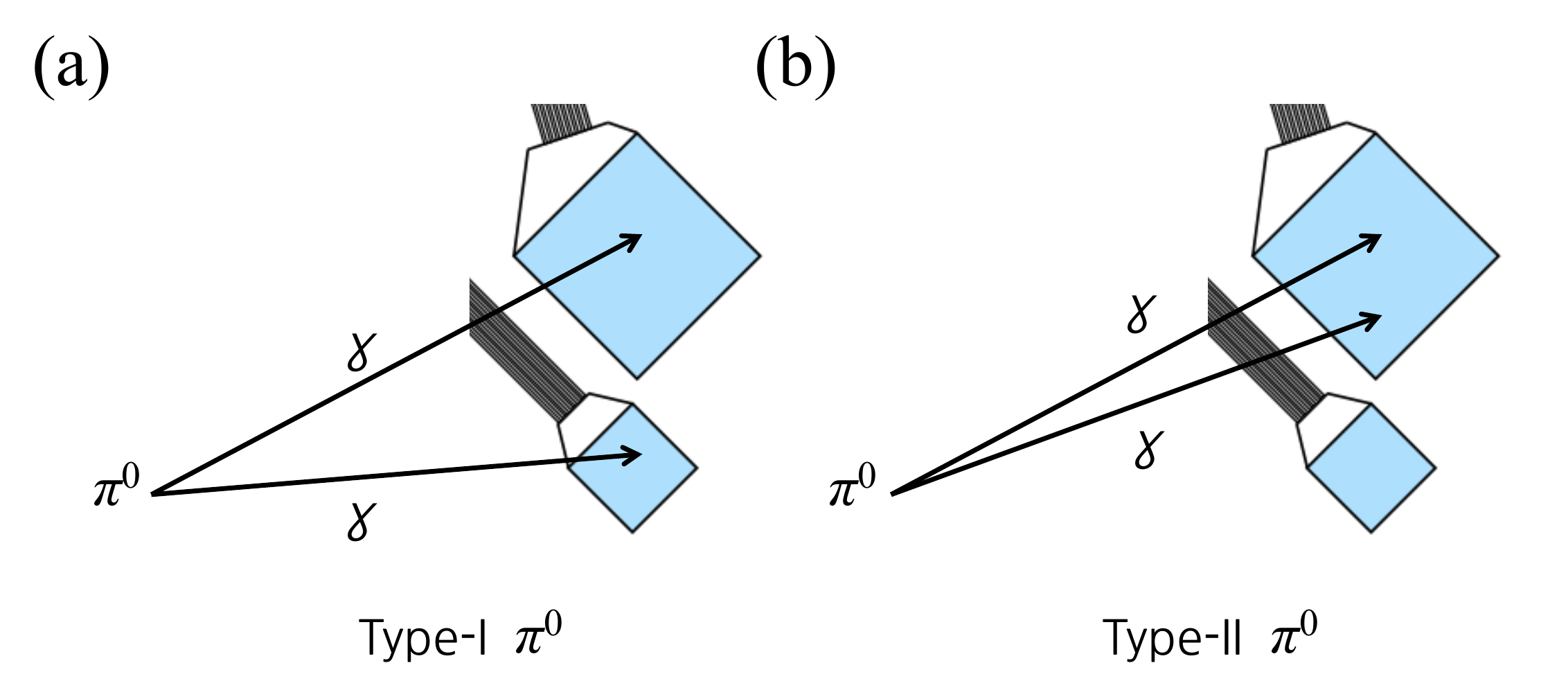}
\caption{Two ways how $\pi^{0}$ is detected by RHICf detector. Two decayed photons can be detected by (a) each tower (Type-I) or (b) one tower (Type-II).} \label{fig:pi0type}
\end{figure}

\begin{figure}[t]
\centering
\sidecaption
\includegraphics[width=0.47\textwidth]{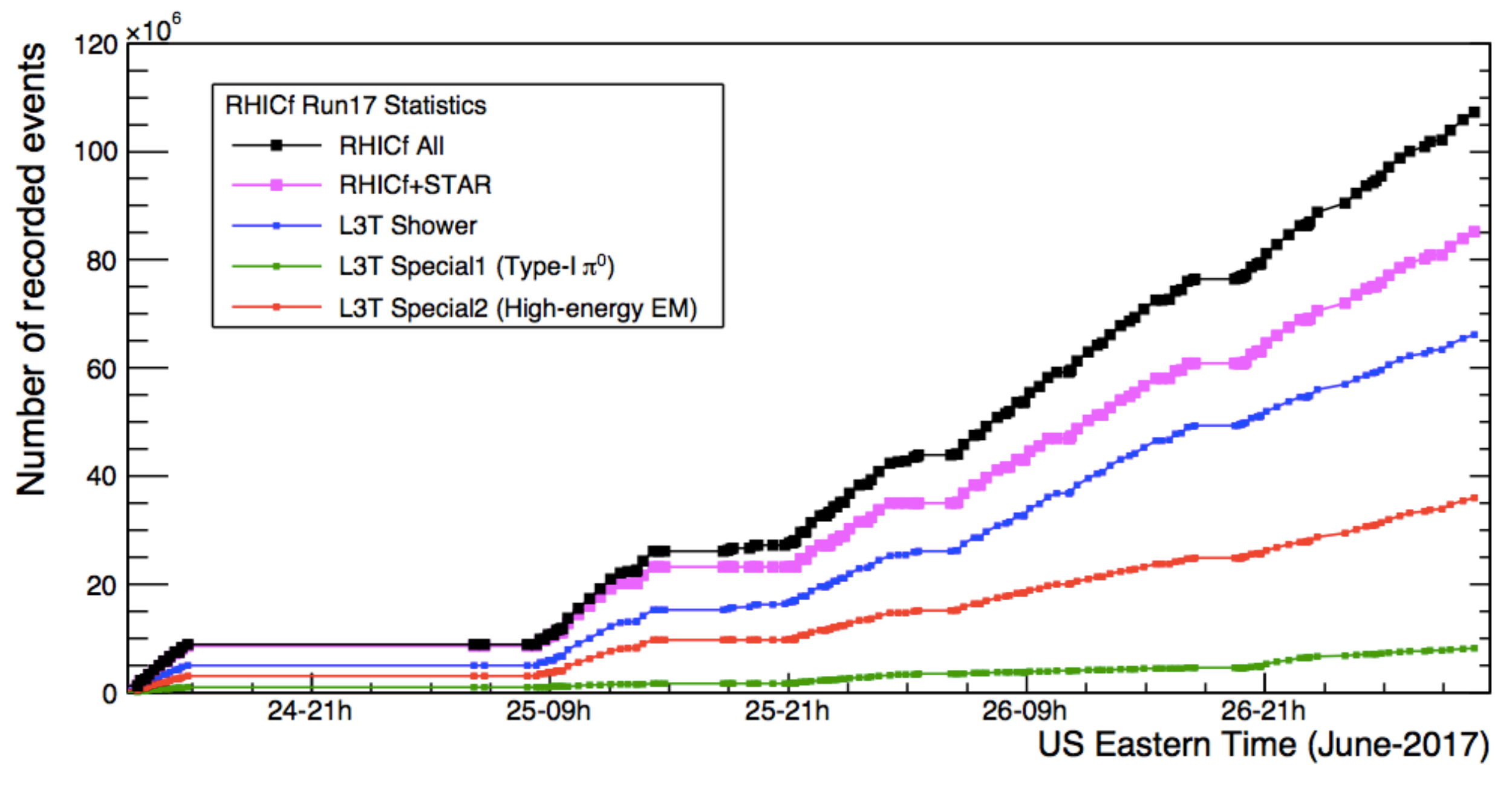}
\caption{Accumulated number of events by each trigger during operation.} \label{fig:events}
\end{figure}

As shown in Fig. \ref{fig:pi0type}, $\pi^{0}$ was measured by two ways. One detecting two decay photons with each tower is called Type-I $\pi^{0}$, and the other detecting them with one tower is called Type-II $\pi^{0}$.
Three kinds of triggers were used to measure the very forward particles. First, a shower trigger was used to measure both single neutrons and photons. It was operated when the energy deposits of three successive layers of small or large towers were larger than 45 MeV. However, because  single photons with lower energy usually dominate the trigger rate and limit the data acquisition live time,  additionally a high energy electromagnetic trigger was operated for the measurement of single higher energy photons and Type-II $\pi^{0}$s. It was defined when the energy deposit of the fourth GSO layer of small or large tower was larger than 500 MeV. Lastly, Type-I $\pi^{0}$ was measured by Type-I $\pi^{0}$ trigger which was operated when the energy deposits of forward three successive layers up to 7$^{th}$ were larger than 45 MeV on both small and large towers. Fig. \ref{fig:events} shows the  accumulated number of events for each trigger during operation. We also recorded the data of the STAR subdetectors according to the RHICf trigger for future combined analysis.

\section{Data analysis}
The $\pi^{0}$ is reconstructed from the two decayed photons. Once the position of a photon was determined by GSO bar layers, its incident energy was reconstructed by position-dependent corrections.
\label{sec:analysis}
\subsection{Position reconstruction}
\label{subsec:positionrec}
If an electromagnetic shower is developed inside the detector by a photon, different energy is deposited in each GSO bar depending on the detector position where the photon hits. The position of a photon is determined by fitting the energy deposit distribution of the layer where the largest energy is deposited. This process is usually done by the 1$^{st}$ or 2$^{nd}$ GSO bar layer due to the radiation length of the detector. 
\subsection{Energy reconstruction}
\label{subsec:energyrec}
Energies deposited in the GSO plates are summed up and the sum energy is corrected into the incident photon energy. Effects of shower particle leakage from the side of the calorimeter and the nonuniform light collection efficiency are corrected layer by layer using the incident position discussed in Sec. \ref{subsec:positionrec}. These conversion function to the incident photon energy and correction functions are determined using a MC simulation taking into account the measured light collection efficiency.
\subsection{$\pi^{0}$ reconstruction}
$\pi^{0}$ can be identified by reconstructing the invariant mass of photon pair events and by extracting events with mass around 135 MeV. For both Type-I and Type-II $\pi^{0}$, the reconstructed energy resolution is around 2.7$\%$ at $\pi^{0}$ energy of 200 GeV and the $p_{T}$ range of 0.2 $\sim$ 0.4 GeV/$c$. The reconstructed $p_{T}$ resolution of Type-I $\pi^{0}$ (0.005 GeV/$c$) is much better than Type-II (0.018 GeV/$c$) because of the difficulty in the Type-II analysis that requires identification of two photons in a single tower.

\begin{figure}[t]
\centering
\sidecaption
\includegraphics[width=0.35\textwidth]{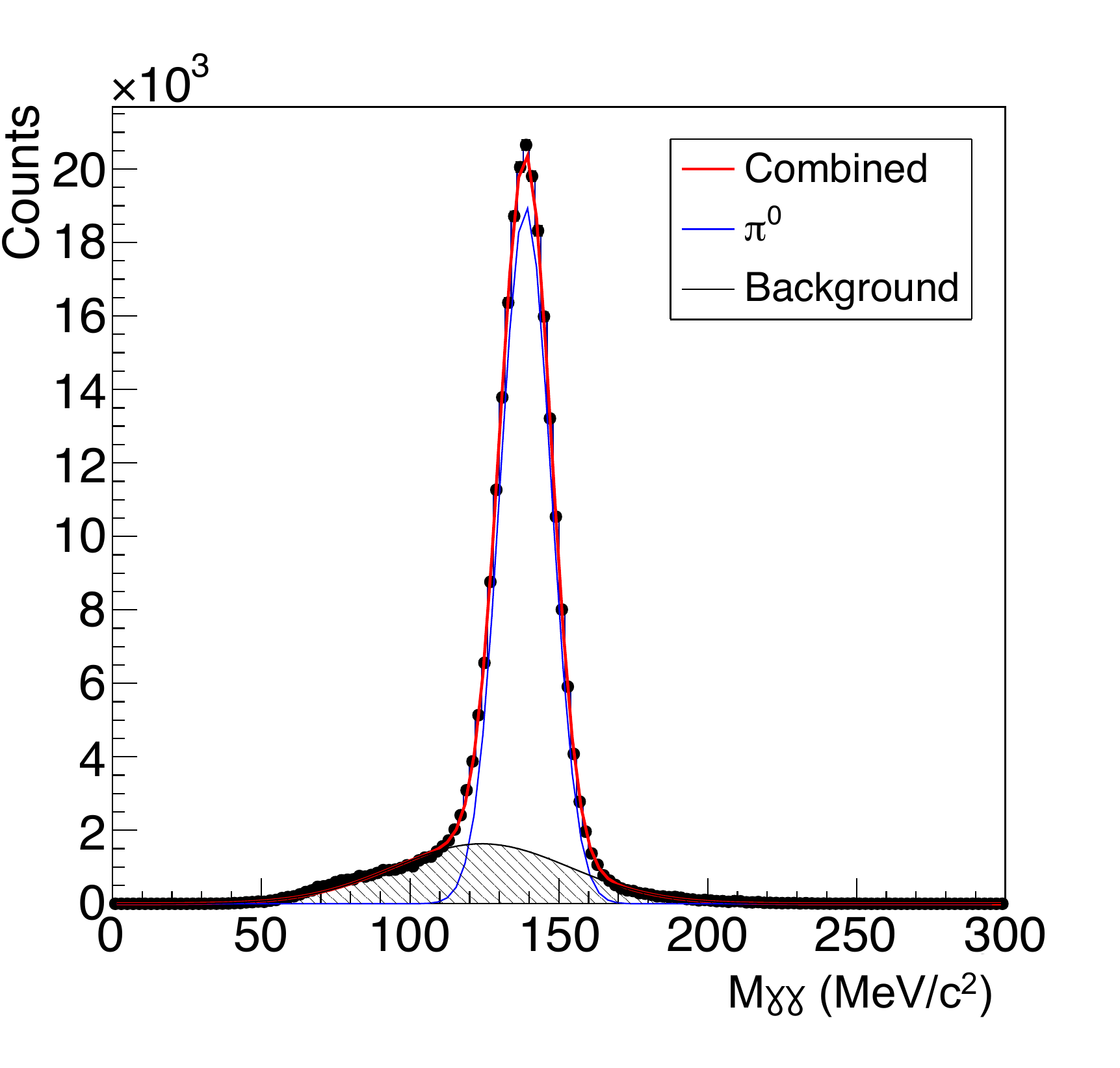}
\caption{Reconstructed invariant mass of two photon events. Black filled area is the background region estimated by fitting.} \label{fig:mgg}
\end{figure}

\begin{figure}[ht]
\centering
\sidecaption
\includegraphics[width=0.4\textwidth]{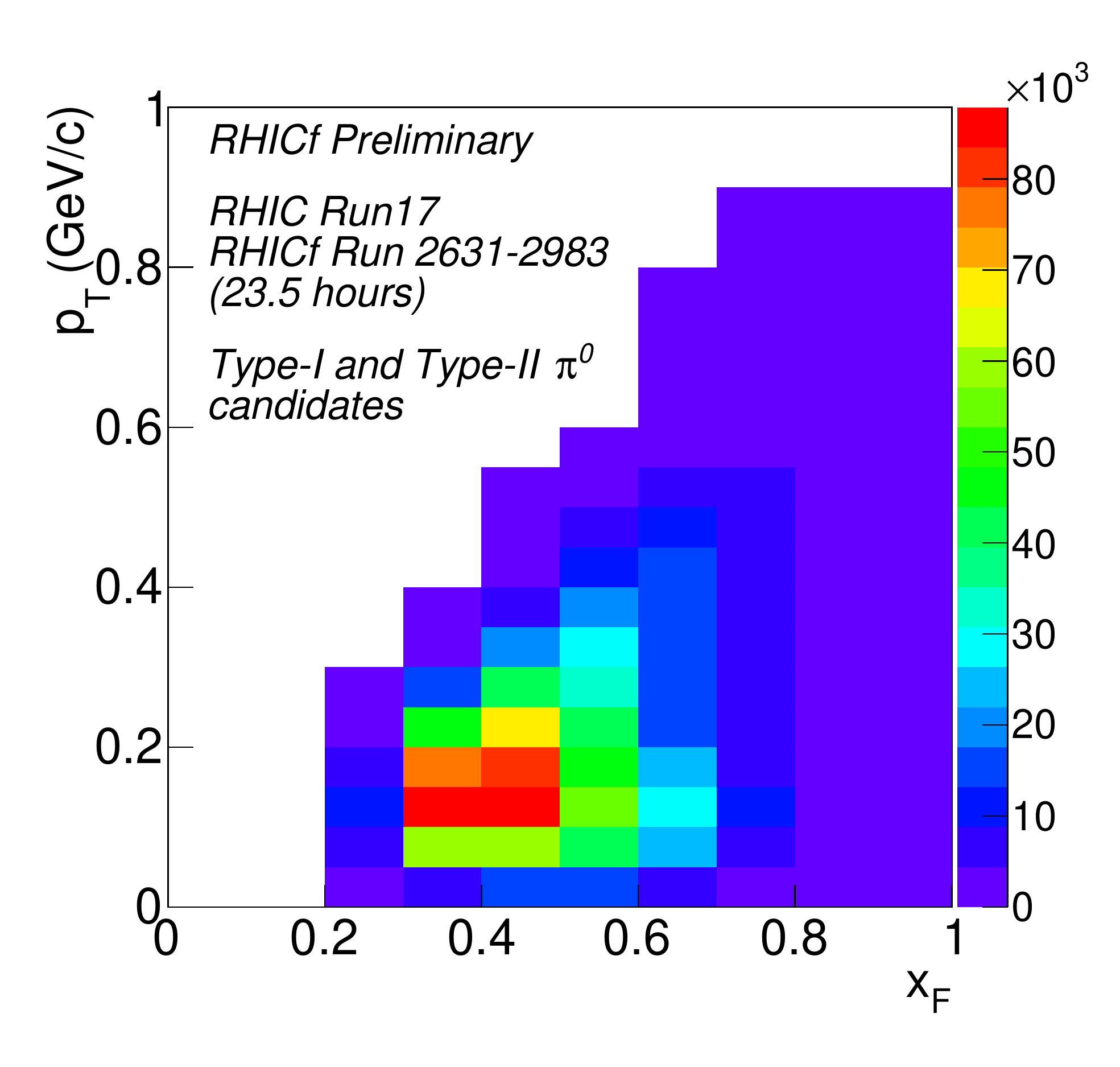}
\caption{Number of final $\pi^{0}$ candidate events as a function of $x_{F}$ and $p_{T}$.} \label{fig:xFpT2}
\end{figure}

The distribution of the reconstructed invariant mass of two photon events is shown in Fig. \ref{fig:mgg}.
A clear $\pi^{0}$ peak is found around 135 MeV/$c^{2}$ with 10 MeV/$c^{2}$ peak width. It was fitted by superposition of polynomial function for background region and Gaussian function for actual $\pi^{0}$. The final $\pi^{0}$ candidate was chosen in $\pm$3$\sigma$ tolerance from the estimated $\pi^{0}$ peak. Fig. \ref{fig:xFpT2} shows the number of $\pi^{0}$ candidate events as a function of $x_{F}$ and $p_{T}$. Both Type-I and Type-II $\pi^{0}$ are included. We first studied the $A_{N}$ in very forward $\pi^{0}$ production for three different Feynman  $x$ ($x_{F}$) ranges for every 0.1 GeV $p_{T}$ bin. Note that the energy and $p_{T}$ resolution of $\pi^{0}$ is much better than their binning scale.
\subsection{$A_{N}$ calculation}
$A_{N}$ is calculated by the following formula, 
\begin{eqnarray}
A_{N} = \frac{1}{P}\frac{1}{D_{\phi}}\Big(\frac{N^{\uparrow}-RN^{\downarrow}}{N^{\uparrow}+RN^{\downarrow}}\Big)
\end{eqnarray}
where $P$ is the polarization magnitude of the proton beam and $D_{\phi}$ is the dilution factor to correct the diluted $A_{N}$ by the azimuthal angle distribution of $\pi^{0}$. 
For $\pi^{0}$s produced with larger azimuthal angles to the beam polarization direction, the measured $A_{N}$ is smaller.
Usual RHIC polarization is around $55\%$ and $D_{\phi}$ of our $\pi^{0}$ measurement is typically larger than 0.9. $N^{\uparrow(\downarrow)}$ represents the detected number of $\pi^{0}$ with polarized up (down) proton beam. $R$ corresponds to the luminosity ratio between polarized up and down proton beam. During RHICf operation, the calculated $R$ was always larger than 0.95. By definition, the $A_{N}$ calculation can be free from detection efficiency effects.  

\section{Result}
\label{sec:result}
\begin{figure}
\centering
\sidecaption
\includegraphics[width=0.4\textwidth]{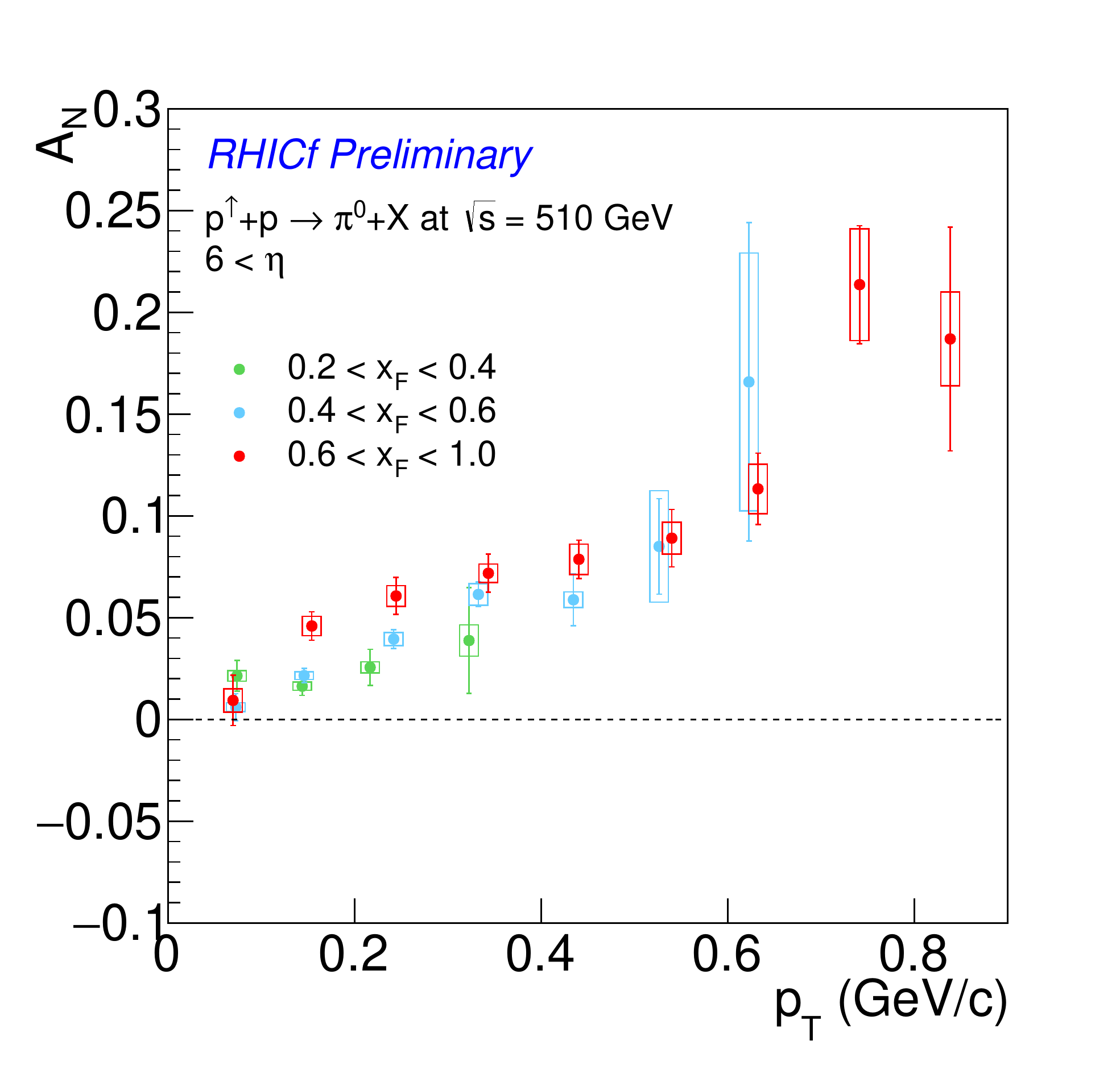} 
\caption{$A_{N}$ in very forward $\pi^{0}$ production as a function of $p_{T}$. Data points with different colors correspond with different $x_{F}$ ranges. Error bars show statistical uncertainties and error bands show systematic uncertainties.} \label{fig:result}
\end{figure}

The preliminary result of the $A_{N}$ in very forward $\pi^{0}$ production is presented in Fig. \ref{fig:result}.
Systematic uncertainties shown by error bands include beam polarization, dilution factor, beam center calculation, and background $A_{N}$ subtraction. Surprisingly, non-zero $A_{N}$ was also observed in very forward $\pi^{0}$ production as well and its magnitude at highest $p_{T}$ is even higher than previous measurements for forward $\pi^{0}$ production at the PHENIX and STAR experiments (See Ref. \cite{phenix}.). Our preliminary result suggests a possible contribution of the diffractive process to the $A_{N}$ of $\pi^{0}$. Because this is an inclusive measurement, we will apply event selections using other STAR detectors to widely study the role of the diffractive and non-diffractive interaction to the non-zero $A_{N}$ of $\pi^{0}$.

\section{Summary}
\label{sec:summary}
The RHICf experiment measured very forward ($6 < \eta$) neutral particles in June 2017 at the interaction point of the STAR experiment. Recently, a  preliminary result of the $A_{N}$ in very forward $\pi^{0}$ production is released and it shows higher $A_{N}$ than previous measurement though the $x_{F}$ and $p_{T}$ ranges are different. For the detailed study to understand the origin of the non-zero $A_{N}$, further analysis will be performed with other STAR detectors.\\[1ex]

\section*{Acknowledgements}  
The RHICf members thank the BNL staff, STAR and PHENIX Collaborations for their essential contributions to the successful operation of RHICf. This work was partially supported by the U.S.-Japan Science and Technology Cooperation Program 
in High Energy Physics and the National Research Foundation of Korea (No. 2016R1A2B2008505).

\end{document}